\documentclass[showpacs,showkeys,prb,twocolumn,aps,preprintnumbers,amsmath,amssymb]{revtex4-2}
\usepackage[breaklinks=true,colorlinks,citecolor=blue,linkcolor=blue,urlcolor=blue]{hyperref}
\usepackage{graphicx,dcolumn,bm,multirow,hyperref,mathtools,braket,color,url,epstopdf,amssymb,amsmath}
\usepackage{amsfonts}

\begin{document}

\title{Analysis of the unconventional chiral fermions in a \\ non-centrosymmetric chiral crystal $\textbf {PtAl}$}

\author{Vikas Saini}
\email{vikas.saini@tifr.res.in}
\affiliation{Department of Condensed Matter Physics and Materials Science, Tata Institute of Fundamental Research, Mumbai 400005, India}

\author{Souvik Sasmal}
\affiliation{Department of Condensed Matter Physics and Materials Science, Tata Institute of Fundamental Research, Mumbai 400005, India}

\author{Ruta Kulkarni}
\affiliation{Department of Condensed Matter Physics and Materials Science, Tata Institute of Fundamental Research, Mumbai 400005, India}

\author{Bahadur Singh}
\email{bahadur.singh@tifr.res.in}
\affiliation{Department of Condensed Matter Physics and Materials Science, Tata Institute of Fundamental Research, Mumbai 400005, India}

\author{A. Thamizhavel}
\email{thamizh@tifr.res.in} 
\affiliation{Department of Condensed Matter Physics and Materials Science, Tata Institute of Fundamental Research, Mumbai 400005, India}

\date{\today}

\begin{abstract}

Symmetry-protected non-trivial states in chiral topological materials hold immense potential for fundamental science and technological advances. Here, we report electrical transport, quantum oscillations, and electronic structure results of a single crystal of chiral quantum material $\rm PtAl$. Based on the de Haas-van Alphen (dHvA) oscillations, we show that the smallest Fermi pocket ($\alpha$) possesses a non-trivial Berry phase $1.16$$\pi$. The band associated with this Fermi pocket carries a linear energy dispersion over a substantial energy window of $\sim$700 meV that is further consistent with the calculated optical conductivity. First-principles calculations unfold that $\rm PtAl$ is a higher-fold chiral fermion semimetal where structural chirality drives the chiral fermions to lie at the high-symmetry $\Gamma$ and $R$ points of the cubic Brillouin zone. In the absence of spin-orbit coupling, the band crossings at $\Gamma$ and $\rm R$ points are three- and four-fold degenerate with a chiral charge of $-2$ and $+2$, respectively. The inclusion of spin-orbit coupling transforms these crossing points into four- and six-fold degenerate points with a chiral charge of $-4$ and $+4$. Nontrivial surface states on the $(001)$ plane connect the bulk projected chiral points through the long helical Fermi arcs that spread over the entire Brillouin zone.    
      
\end{abstract}

\maketitle

\section{INTRODUCTION}
The interplay between crystalline symmetries and band topology drives exotic states in materials, which have been investigated intensively over the years for their attributes to experimental standouts~\cite{vergniory2019complete,turner2013beyond,zhang2019catalogue,cano2019multifold}. Dirac and Weyl fermions predicted as the fundamental particles in high-energy physics have been reported to exist as low-energy excitations in topological quantum materials~\cite{bradlyn2016beyond,liu2014discovery,kushwaha2015bulk,wang2013three}. Four-fold degenerate nodes in Dirac semimetals are theoretically predicted and subsequently realized in experiments. The energy dispersion near Dirac nodes is linear in all three spatial directions that mimic the Dirac equation with a net-zero chiral charge. These Dirac points are protected by a combination of inversion, time-reversal, and lattice symmetries. By breaking either inversion or time-reversal, a Dirac node splits into two opposite charged Weyl points where energy dispersion is described by the chiral Weyl equation~\cite{lv2015observation, sun2015topological, xu2015experimental,LaAlGe}. Chirality, degeneracy, and dimensionality of the crossing points distinguish numerous topological states in materials~\cite{chiu2016classification,senthil2015symmetry,BaAgAs}. 
Non-symmorphic symmetries can impose band crossings at high symmetry points in the Brillouin zone (BZ) and produce higher-fold fermions without any high-energy analogue. Specifically, cubic systems of space group $P2_13$ ($\#198)$ such as $\rm Rh(Si, Ge)$, $\rm (Pd, Pt)Ga$ etc. evince non-symmorphic symmetry protected multifold fermions at the high symmetry points with unconventional low-energy excitations called spin-$\frac{3}{2}$ Rarita-Schwinger fermions ~\cite{rees2020helicity,chang2017unconventional,sessi2020handedness, yao2020observation, li2019chiral, gao2019topological,chang2018topological}. The multifold chiral fermions produce novel optoelectronic responses such as the photogalvonic effect~\cite{de2017quantized,flicker2018chiral, sanchez2019linear, wu2017giant, ma2017direct, ni2020linear, xu2020optical}.  In this work, we have investigated the topological quantum properties of single crystals of chiral material $\rm PtAl$ combining theoretical analysis and experimental magnetotransport measurements. Hall resistivity measurements revealed that the electrons dominate over hole carriers in the field range $0-14$~T and in the temperature range $1.8 - 100$~K. The Berry phase is estimated from the de Haas-van Alphen (dHvA) oscillations. Linear $\alpha$ and parabolic $\beta$ bands are observed at $\Gamma$ high symmetry point possessing non-trivial and trivial Berry phases, respectively.  The first-principles calculations reveal the three- and four-fold nodal points at $\Gamma$ and $R$ time-reversal invariant momentum (TRIM) points near the Fermi level without spin-orbit coupling.  In the presence of spin-orbit coupling,  the nodal crossings at $\Gamma$ and $R$ points evolve into the four-fold and six-fold degenerate nodal point containing low energy excitations as the Rarita-Schwinger fermions. We also calculate the optical response, which reveals that optical conductivity is almost linear with frequency in the energy range from $400$ to $800$~meV suggesting the linear band transitions from the near Fermi energy bands. Unconventional electronic and optical properties show that PtAl is an interesting chiral material for investigating topological properties and device applications.

\section{METHODS}

{\it Single crystal growth--}
The congruent melting nature of PtAl facilitated us to grow the single crystal by the Czochralski method in a tetra-arc furnace.  High purity starting materials of Pt (3N pure) and Al (5N pure) were taken in the stoichiometric ratio, with a little excess of Al to compensate for the weight loss of Al evaporation during the growth. The phase purity of the grown crystal is confirmed by powder x-ray diffraction (XRD) studies performed at room temperature using a Rigaku x-ray diffractometer and the orientation of the crystal was done using Laue diffraction in the back-reflection geometry.  The electrical resistivity measurements are carried out on a flat shiny surface of  0.6$\times$1.32$\times$0.055~mm$^{3}$ dimensional crystal, and contacts are made with a two-component Ag paste using $50$~$\mu$m diameter gold wires in a physical property measurement system (PPMS), Quantum Design. Magnetization measurements are done in a vibrating sample magnetometer equipped with a $14$~T magnet.

{\it Computational details --}
Electronic structure calculations were performed within the density functional theory (DFT) framework using the Quantum Espresso software package~\cite{giannozzi21others,giannozzi2020quantum}. A 21$\times$21$\times$21 $k$-mesh was used for the BZ sampling, and a kinetic energy cut-off of 60 Ry was used for the plane-wave basis set. The optical properties were calculated using the full potential linearized augmented plane-wave (FL-LAPW) method as implemented in the WIEN2k package~\cite{blaha2001wien2k,schwarz2003solid,schwarz2003dft}. A denser $k$-mesh of 46$\times$46$\times$46 was used to obtain the well converged results. We constructed the first-principles tight-binding model Hamiltonian and calculated the topological properties using the WANNIERTOOLS~\cite{mostofi2008wannier90,kunevs2010wien2wannier,WU2017}.


\section{RESULTS AND DISCUSSIONS}

\subsection{Crystal Structure and X-ray Diffraction Measurements}

\begin{figure}[h]
\includegraphics[width=0.5\textwidth]{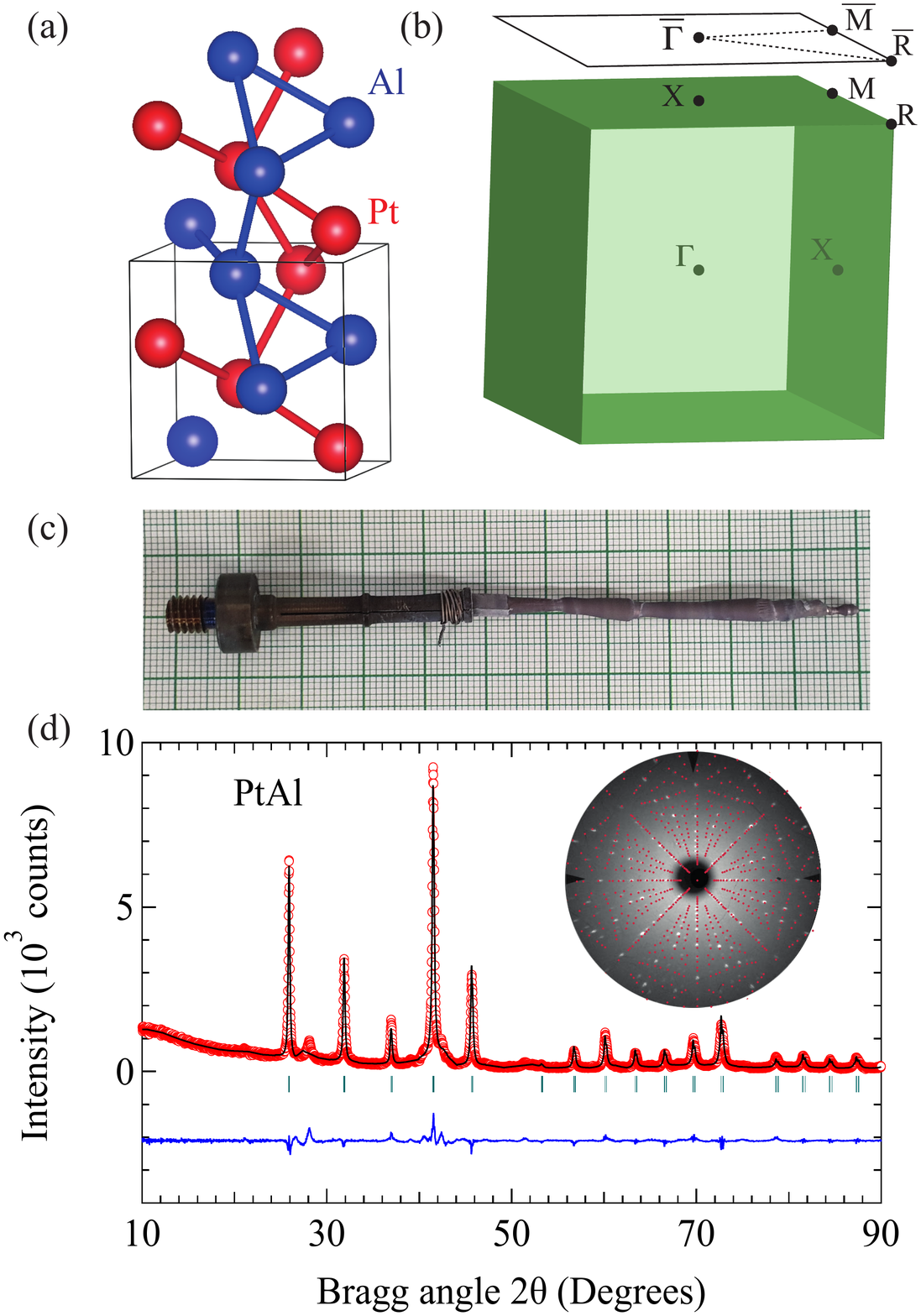}
\caption{(a) (a) The cubic crystal lattice of $\rm PtAl$ with the chiral structure. (b) The associated bulk and (001) surface Brillouin zones. The high-symmetry TRIM points are marked. (c) As grown single-crystal of PtAl. (d) Powder XRD pattern of PtAl along with the Rietveld refinement. The inset shows the observed Laue diffraction pattern overlapped with the simulated pattern corresponding to the (100)-plane.}
	\label{Fig1}
\end{figure}

PtAl crystallizes in the noncentrosymmetric cubic space group P$2_1$3$(\#198)$, as shown in Fig.~\ref{Fig1} (a).  Since the space group includes the screw-axis symmetry operation it falls in the non-symmorphic chiral space group category.  Figure~\ref{Fig1}(b) depicts the associated BZ where the high symmetry TRIM points are marked. As grown pulled ingot of single crystal of PtAl is shown in Fig.~\ref{Fig1}(c). We confirmed the phase purity of the single crystal by grinding a small portion of the crystal to fine powders and by performing the powder XRD using the monochromatic Cu-$k_{\rm \alpha}$ wavelength x-ray source. The Rietveld analysis of the powder XRD using the FullProf software is shown in Fig.~\ref{Fig1}(d). The lattice constant was extracted as $a~=4.864$~\AA, which is used in our first-principles computations. The inset in Fig.~\ref{Fig1}(d) shows the observed Laue diffraction pattern together with the simulated pattern corresponding to the (100) plane. Well defined diffraction spots with $4$-fold symmetry confirm the good quality of our grown single crystals. This crystal was cut along the [100] direction for the electrical resistivity and dHvA studies whereas the plane normal to this direction is (110).   

\subsection{Resistivity Measurements}
\begin{figure}[!]
\includegraphics[width=0.5\textwidth]{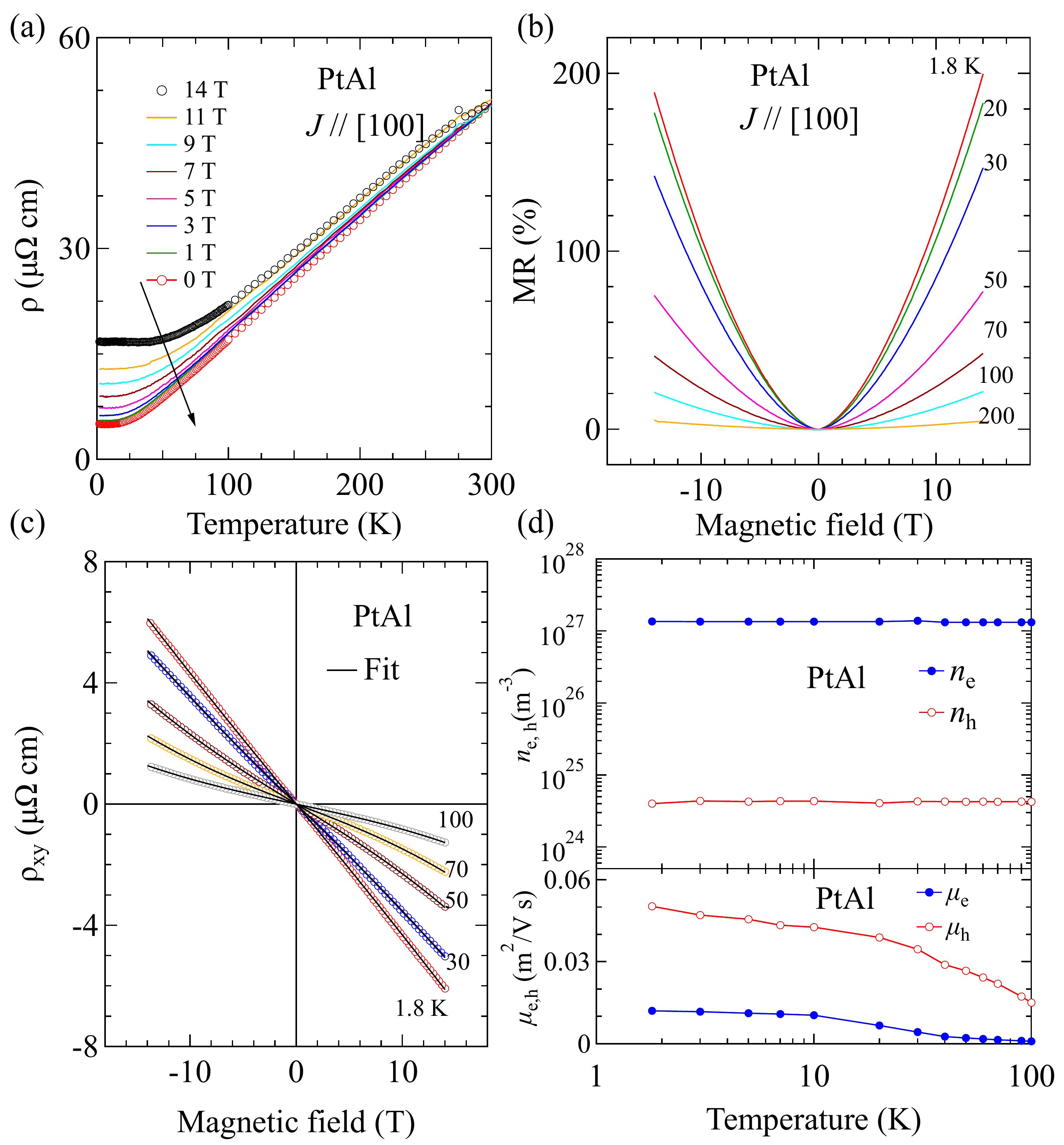}
\caption{Electric transport measurements of PtAl. (a) Resistivity as a function of temperature (1.8 to 300 K) for various magnetic fields. (b) Magnetoresistance vs applied magnetic field of $-14$ to $14$~T at several temperatures. (c) Hall resistivity vs applied magnetic field in the field range $-14$ to $+ 14$~T, measured at various constant temperatures where solid line is corresponding to two-band model fitting. (d) Carrier densities and mobilities as a function of temperature.}\label{Fig2}
\end{figure}

The electrical resistivity measured for applied magnetic fields in the temperature range from 2 to 300~K for $J~\parallel$~[100] is shown in Fig.~\ref{Fig2}(a). The resistivity decreases as the temperature is decreased from 300~K and follows a nearly linear dependence down to 50~K and a change of slope is observed at around 30~K and remains almost constant at around $5$~$\mu \Omega$ cm for temperature less than $10$~K. At $300$~K the resistivity attains a value of $51$~$\mu \Omega$ cm, and hence the residual resistivity ratio (RRR) is about 10. With the application of a magnetic field, the electrical resistivity increases at low temperatures. The magnetoresistance (MR) is shown in Fig.~\ref{Fig2}(b) in the field range $-14$ to $+14$~T, for various temperatures. At $2$~K, the MR attains a value of 195\% and gradually drops to lower values with increasing temperature.  

The magnetic field dependence of Hall resistivity $\rho_{\rm xy}$ measured at various temperatures is shown in Fig.~\ref{Fig2}(c). The Hall data was antisymmetrized ($\rho_{xy}(B)$ $=$ $\frac{\rho_{xy}(B)-\rho_{xy}(-B)}{2}$) to avoid the longitudinal resistivity contribution from the possible misaligned electrical contacts on the sample surface. In the low-temperature range from $1.8 -10$~K, the Hall resistivity is linear with a negative slope suggesting electrons are the dominant carriers. However, non-linearity in the Hall resistivity is observed in high-temperature, above $10$~K (Fig~\ref{Fig2}(c)). The carrier concentrations and the mobilities are obtained by employing the two-band model:

\begin{equation}
	\rho_{xy} = \frac{B}{e} \frac{(n_{h} \mu_{h}^{2}-n_{e} \mu_{e}^{2}) + (n_{h}-n_{e})(\mu_{h}\mu_{e})^2 B^2}{(n_{h} \mu_{h}+n_{e} \mu_{e})^2+(n_{h}-n_{e})^2(\mu_{h}\mu_{e})^2 B^2 },
	\label{Eq1}
\end{equation} 

where $n_e$, $n_h$ are the carrier concentration of the electrons and holes respectively and $\mu_e$, $\mu_h$ are the mobilities of electrons and holes, respectively. Two-band model fitted Hall resistivity data is shown in Fig~\ref{Fig2}(c). The estimated carrier concentrations and the mobilities from the fitting are plotted in Fig~\ref{Fig2}(d). The carrier densities are almost temperature-independent and the electron concentration is $\sim$~300 times more than the hole density. In addition, the hole mobilities are found to be greater than the electron mobilities in the entire range of measurement. These results are consistent with our first-principles results as discussed in Sec.~\ref{el_str}.

\subsection{de Haas-van Alphen oscillations}

\begin{figure*}[!]
\includegraphics[width=1.0\textwidth]{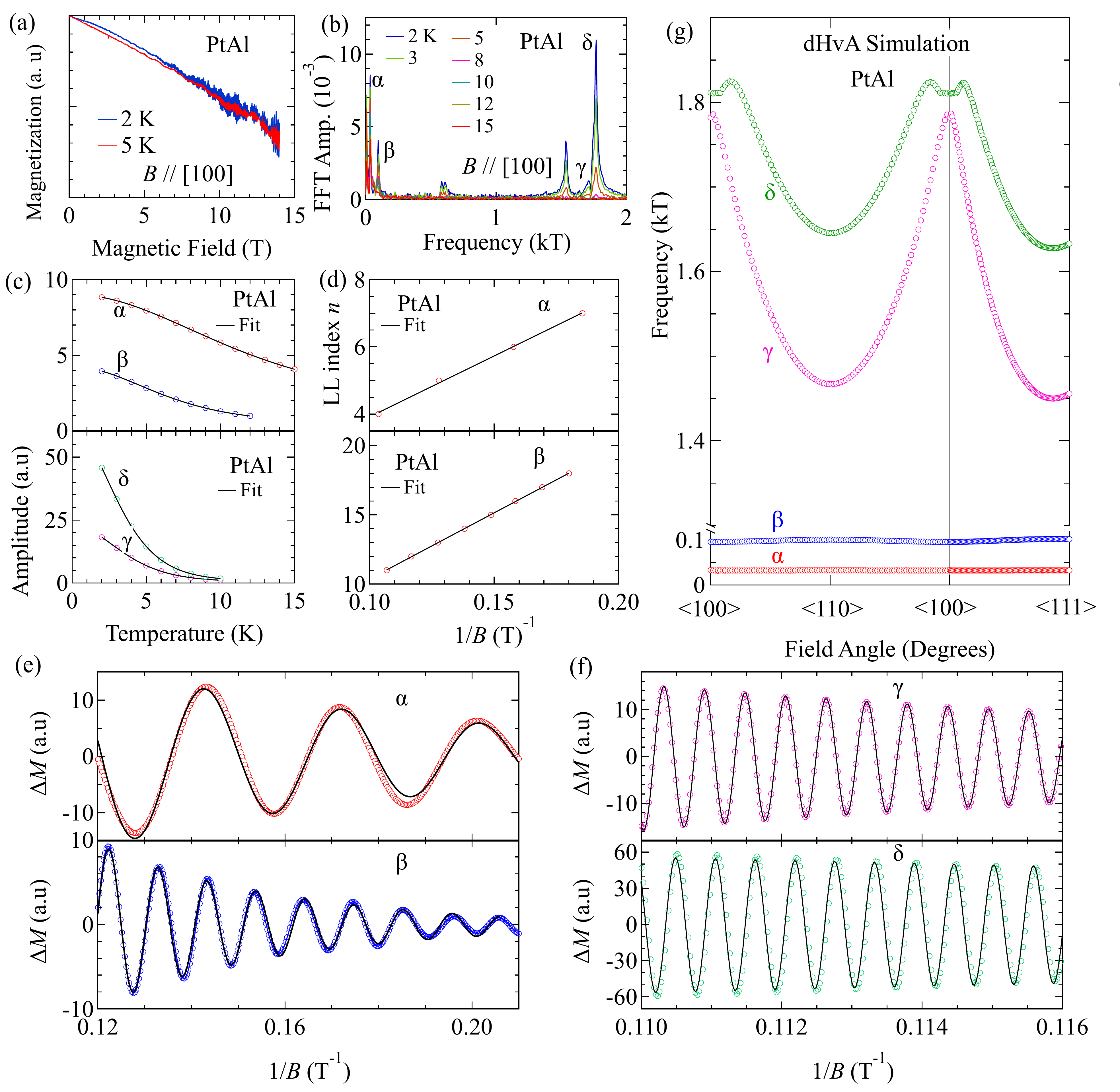}
\caption{(a) De Haas-van Alphen (dHvA) oscillations observed at $2$ and $5$~K for $B~\parallel$~[100]. (b) FFT spectra of the dHvA oscillations measured at various temperatures. (c) Temperature dependence of the FFT amplitudes. The solid lines are the fit to the thermal damping term ($R_{T}$) of the L-K formula. (d) Landau fan diagram plotted against an inverse magnetic field where black solid line is fit to the data points. (e, f) Band pass filtered quantum oscillations where the solid line is fit of L-K equation at temperature 2 K. (g) The simulation of quantum oscillations for $\alpha$, $\beta$, $\gamma$, and $\delta$ bands for various orientations of the magnetic field.}\label{Fig3}
\end{figure*}

To probe the Fermi surface properties of PtAl, we performed the de Haas-van Alphen (dHvA) oscillation studies by measuring the isothermal magnetization along [100] direction by ramping the magnetic field from 0 to $14$~T. The representative plots of isothermal magnetization measured at $2$ and $5$~K, depicting the dHvA quantum oscillations, are shown in Fig.~\ref{Fig3}(a). Oscillations appear for fields greater than $4$~T, and they are superimposed with the diamagnetic signal. The background is carefully subtracted and the oscillation frequencies are extracted by performing the fast Fourier transformation (FFT) on the background-subtracted data as shown in Fig.~\ref{Fig3}(b). Four fundamental frequencies are observed at $36$, $95$, $1721$, and $1764$~T that corresponds to $\alpha$ , $\beta$, $\gamma$, and $\delta$ bands in the bulk band structure calculations. The frequency spectra of the dHvA quantum oscillations are analyzed to extract the quantum parameters from the well known Lifshitz-Kosevich formula given in Ref.~\cite{saini2021fermi},

\begin{table*}
\caption{\label{Tab1}Summary of estimated quantum parameters from dHvA oscillations for $\alpha$, $\beta$, $\gamma$, and $\delta$ bands. $F_{exp}$ and $F_{cal}$ are corresponding to experimentally observed and calculated dHvA frequencies, respectively. $m^*_{exp}$ and $m^*_{cal}$ are the experimental and calculated effective masses. In addition, other quantum parameters obtained as; Dingle temperature $T_{D}$, quantum relaxation time $\tau_q$, quantum mobility $\mu_q$, Fermi vector $k_{\rm F}$, Fermi velocity $v_{\rm F}$, Fermi level $E_{\rm F}$, mean free path $l_q$, and carrier density $n$.} 
	\begin{ruledtabular}
		\begin{tabular}{cccccccccccccc}
			
			Band & $F_{exp}$	& $F_{cal}$ & $m^{*}_{exp}$ & $m^{*}_{cal}$ & $T_{D}$ & $\tau_{q}$	& $\mu_{q}$ & $k_{\mathrm F} $	& $v_{\mathrm F}$	 & $E_{\rm F}$ & $l_{q}$ & $n$ \\  \\ 
			B$\parallel$[100] & (T)    & (T)     &  $(m_e)$   & $(m_e)$  & (K)   & $(10^{-13}sec)$  &   $(cm^2V^{-1}s^{-1})$   & (\AA$^{-1})$   & $(10^{5}$m/s)   &{(meV)} &   (nm)      &    ($10^{18} cm^{-3}$)  \\ 
			\hline \\
			$\alpha$ & 36  & 33 & 0.1 & 0.06 & 5.70 & 2.13  & 3698     &   0.033   &   3.79 & 82.8   &  80.8   &   1.24 \\ \\
			\hline \\
			
			$\beta$ & 95  & 96 & 0.179(5) & 0.125 & 7.75 &  1.57& 1544& 0.054 & 3.47 & 421 & 122.6& 5.21\\ \\
			\hline \\

			$\gamma$  & 1721  & 1782 &  0.308(5) & 0.416 & 16.53  &  0.73 & 420 & 0.229 & 8.6 &  1294 & 63.2 &400 \\ \\
			\hline \\
			
			$\delta$ &1764  & 1812 & 0.341(5)  & 0.41 & 3.40 &  3.58 & 1846 & 0.232 & 7.86 & 1197  &281.3 &420 \\  
			\\

		\end{tabular}
	\end{ruledtabular}
\end{table*}

\begin{equation}
	\label{Eqn2}
	\Delta M \propto -B^{k}R_{T}R_{D}R_{s}sin\left[2\pi\left(\frac{F}{B}+\psi \right)\right],
\end{equation}  

where $R_T$, $R_D$, and $R_s$ are the thermal, magnetic field, and spin damping factors and $\psi$ is the phase factor. The effective masses for all the four bands are obtained from the thermal damping factor $R_{T}$ which is given by $R_T = (X/sinh X)$, where $X = (\lambda T m^*/B)$, $\lambda = (2 \pi^2 k_B m_e/e\hbar)$; $m^{*}$ is the effective mass of carriers in units of rest mass of electron.  For $\alpha$ band effective mass appears to be small as 0.1~$m_{e}$ and $\delta$ band has the highest mass among the four bands as 0.341(5)~$m_{e}$, whereas the estimated masses for $\beta$ and $\gamma$ bands are 0.179(5) and 0.308(5)~$m_e$, respectively as estimated from Fig~\ref{Fig3}(c).  The small mass of $\alpha$ band is attributed to the  linear band dispersion at $\Gamma$ high symmetry point (~$\sim$~700 meV), close to the Fermi energy.   To extract the Berry phase, we plot the Landau level fan diagram as shown in Fig.~\ref{Fig3}(e-f).  Here we indexed the minima of dHvA oscillations as ($n- \frac{1}{4}$)$^{\rm th}$ Landau level; thus Berry phase for $\alpha$ and $\beta$ bands are estimated using relation $n - \frac{1}{4}= F/B + \psi $ where $\psi$ = ($\frac{1}{2}$ $-$ $\frac{\Phi_{B}}{2\pi}$) $-$ $\delta$. $\delta$ is 0 for 2D nature of Fermi pocket, and $\pm$ $\frac{1}{8}$ for hole and electron pockets respectively.

To facilitate the direct comparison with our experimental measurements, we simulate the quantum oscillations frequencies and effective masses of $\alpha$, $\beta$, $\gamma$, and $\delta$ bands (see Table~\ref{Tab1}). The experimental frequencies closely match with the calculated frequencies when the Fermi energy level is shifted upward in energy by $\sim$~29 meV, suggesting n-type doping in our crystals~\cite{hsieh2009tunable}. Specifically, $\alpha$ and $\beta$ electron bands are found at the $\Gamma$ point (bands in cyan color) with the linear and parabolic energy-momentum relations, respectively. These bands form small Fermi surfaces as illustrated in Fig.~\ref{Fig4}(b) and Fig.~\ref{Fig4}(h). Whereas $\gamma$ and $\delta$ bands are identified at the R point and represented in blue color in the band structure (Fig.~\ref{Fig4}(b)). The calculated quantum oscillations of $\alpha$, $\beta$, $\gamma$, and $\delta$ bands for various orientations of the magnetic field are shown in Fig.~\ref{Fig3}(g). The oscillations frequencies of $\alpha$ and $\beta$ bands are nearly constant with changing the magnetic field direction which reveal their three-dimensional character. On the other hand,  $\gamma$ and $\delta$ bands are highly anisotropic. $\gamma$ band oscillations frequency is $\sim$~1782 T when the magnetic field is parallel to  $<$100$>$. It shows a minimum at 1467 T and 1456 T for B $\parallel$ $<$110$>$ and B $\parallel$ $<$111$>$ directions. The oscillation associated with the $\delta$ band is found to be $\sim$~1812 T for $<$100$>$ direction. It increases up to certain angles with rotating the magnetic field towards $<$110$>$ and $<$111$>$ directions. Beyond these angles, the oscillations frequency decreases and attains a minimum at 1646 T and 1633 T for  B $\parallel$ $<$110$>$ and  B $\parallel$ $<$111$>$ directions.

Since $\alpha$ and $\beta$ are obtained as electron-type three-dimensional bands, we consider dimensionality factor $\delta$ of -$\frac{1}{8}$ for Berry phase estimation. The phase factors for $\alpha$ and $\beta$ bands obtained from the intercept of the Landau fan diagram (Fig.~\ref{Fig3}(d)) are estimated to be 0.29(3) and 0.82(6). The $\alpha$ and $\beta$ bands possess non-trivial and trivial Berry phases of 1.16$\pi$ and 0.11$\pi$, respectively. At temperature 2 K, background subtracted band pass filtered quantum oscillations are fitted with the 3D Lifshitz-Kosevich expression (given in Eq.~\ref{Eqn2}) for the $\alpha$, $\beta$, $\gamma$, and $\delta$ bands which are shown in Fig.~\ref{Fig3}(e, f). We also estimated the quantum parameters such as quantum scattering time ($\tau_{q}$), quantum mobility ($\mu_{q}$), Fermi vector ($k_{F}$), Fermi velocity ($v_{F}$), Fermi level ($E_{F}$), and quantum mean free path ($l_{q}$) for observed $\alpha$, $\beta$, $\gamma$, and $\delta$ bands and  are summarized in Table~\ref{Tab1}. Since the linear energy-momentum relation of the $\alpha$ band leads to the high mobility; that is consistent with the estimated quantum mobility which shows the highest mobility among all four observed bands as given in Table~\ref{Tab1}. 

\subsection{Electronic Structure}\label{el_str}

\begin{figure}[!]
\includegraphics[width=0.5\textwidth]{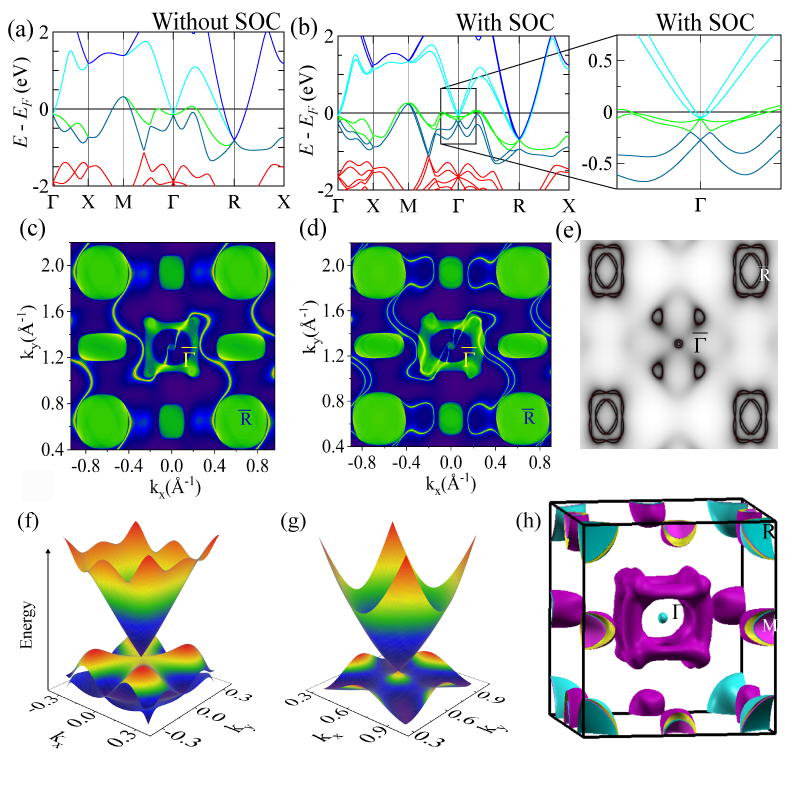}
\caption{Calculated band structure of $\rm PtAl$ (a) without and (b) with spin-orbit coupling along high-symmetry directions in the cubic Brillouin zone.  The inset highlights the closeup of bands at $\Gamma$ in (b). Fermi contours associated with (001) surface spectrum (c) without and (d) with spin-orbit coupling. The Fermi arcs connecting $\Gamma$ points are resolved.  (e) Bulk Fermi surface cut at $k_{z}=0$ plane. Three-dimensional energy dispersion ($E-k_x-k_y$) of the bulk band crossings at (f) $\Gamma$ and (g) $\rm R$ points in the absence of spin-orbit coupling. (h) Calculated bulk Fermi surfaces of PtAl.  }
	\label{Fig4}
\end{figure}

We now discuss the topological electronic properties of PtAl. The calculated bulk band structure without and with spin-orbit coupling (SOC) is shown in Figs.~\ref{Fig4}(a-b). It is metallic with both the electron and hole pockets present at the Fermi level. The valence and conduction bands cross at $\Gamma$ and $\rm R$ points below the Fermi level without SOC in Fig.~\ref{Fig4}(a). Since PtAl lattice is chiral without inversion symmetry, the band crossings at high-symmetry points realize Kramer's Weyl fermions~\cite{chang2017unconventional}. We have further calculated the topological charge that shows that three-fold band crossings lying at $\sim 0.14$~eV below the Fermi level at $\Gamma$ point possess a chiral charge of $-2$. The low-energy physics of these crossings can be described by spin-1 excitations in contrast to $C_{3v}$ symmetry-protected band crossings in triple-fermions semimetals $\rm MoP$, $\rm WC$, $\rm ZrTe$, $\rm BaAgAs$ materials~\cite{BaAgAs,lv2017observation, ma2017three, winkler2019topology}. At the BZ corner ( R point), the crossing is fourfold and lies $\sim 0.8$~eV below the Fermi level. It carries a chiral charge $+2$ and differs from achiral four-fold Dirac points. Two Weyl cones of the same chirality combine to form four-fold degeneracy at the $\rm R $ point. The three-dimensional rendition of band crossings near $\Gamma$ and $\rm R$ points is shown in Fig~\ref{Fig4}(f-g). At $\Gamma$ point, three-fold band crossings of spin-1 excitations are resolved, whereas nearly two-fold degenerate bands cross at $\rm R$ to generate the Weyl node. The band structure with SOC effects is shown in Fig.~\ref{Fig4}(b). Due to the lack of inversion symmetry, all the bands are spin-split except at the time-reversal momentum points where they remain Kramer's degenerate. The band crossings at $\Gamma$ and $\rm R$ evolve to four- and six-fold degenerated points with the chiral charge $\mp$ 4, respectively. It should be noted that the chiral fermion band crossings at $\Gamma$ and $R$ points lie below the Fermi energy level. These crossing bands at the Fermi level form dominant electron pockets. In contrast, the small hole pockets are located at the $M$ points (Fig.~\ref{Fig4}(b)). Since our samples are electron doped, the hole pockets can shrink further, leading to domanint electron carriers consistent with our Hall measurements~\cite{chang2017unconventional, xu2020optical}.        

The (001) surface Fermi contours without and with SOC are shown in Figs.~\ref{Fig4}(c-d). The bulk chiral nodes on the (001) surface are projected at the center $\overline \Gamma$ and corner $\overline R$ points of the surface BZ. These points are connected with the unconventional long helical arcs. Notably, two Fermi arcs connect $\overline \Gamma$ and $\overline R$ points without SOC. They evolve into four Fermi arcs as seen in Fig.~\ref{Fig4}(d). These results are consistent with the calculated bulk topological charge and reveal that PtAl has long Fermi arcs state covering the full surface BZ. Specifically, the separation between opposite chiral points (1.12~\AA$^{-1}$) is substantially higher than observed values for single Weyl fermion semimetals such as $\rm Mo_{x}W_{1-x}Te_{2}$, $\rm (Ta, Nb)(P, As)$, $\rm WTe_{2}$, $\rm LaAlGe$ ~\cite{chang2017unconventional,huang2015weyl,weng2015weyl,chang2016prediction, ruan2016ideal,LaAlGe}. Among the chiral compounds of the same space group (P$2_1$3), CoSi and RhSi materials have smaller lattice parameters than PtAl therefore, the Weyl points separation in these materials is somewhat more than the PtAl~\cite{xu2020optical, rees2020helicity}. Nonetheless, the remarkably higher separation between the chiral points makes the system robust against the disorders. 	

Figures \ref{Fig4}(e, h) resolve the bulk Fermi surface of PtAl. The Fermi surface cut at $k_{z}$ $=$ 0 plane reveals two electron pockets centered at $\Gamma$ point. These are observed in the dHvA oscillations as $\alpha$ and $\beta$ pockets. The smallest pocket $\alpha$ obeys the linear behaviour dominated over the square and cubic terms in energy dispersion over a large energy window of $\sim$~700 meV and possesses a nontrivial Berry phase of $1.16$$\pi$. $\beta$ pocket is bigger as compared to the $\alpha$ pocket with dominated parabolic energy dispersion and tends to possess the trivial Berry phase of $0.11 \pi$. Figure~\ref{Fig4}(h) shows the merged Fermi surface of all eight bands. The splitting of the bulk and surface states suggests strong SOC strength in $\rm PtAl$ as probed for isostructural compounds $\rm (Pt, Pd)Ga$~\cite{yao2020observation, schroter2020observation}. The effective masses for $\alpha$ (0.1 $m_e$) and $\beta$ (0.179 $\pm$ 0.005 $m_e$) bands observed here are also comparable to that observed for $\rm PtGa$ ~\cite{yao2020observation}.    

\begin{figure}[!]
	\includegraphics[width=0.45\textwidth]{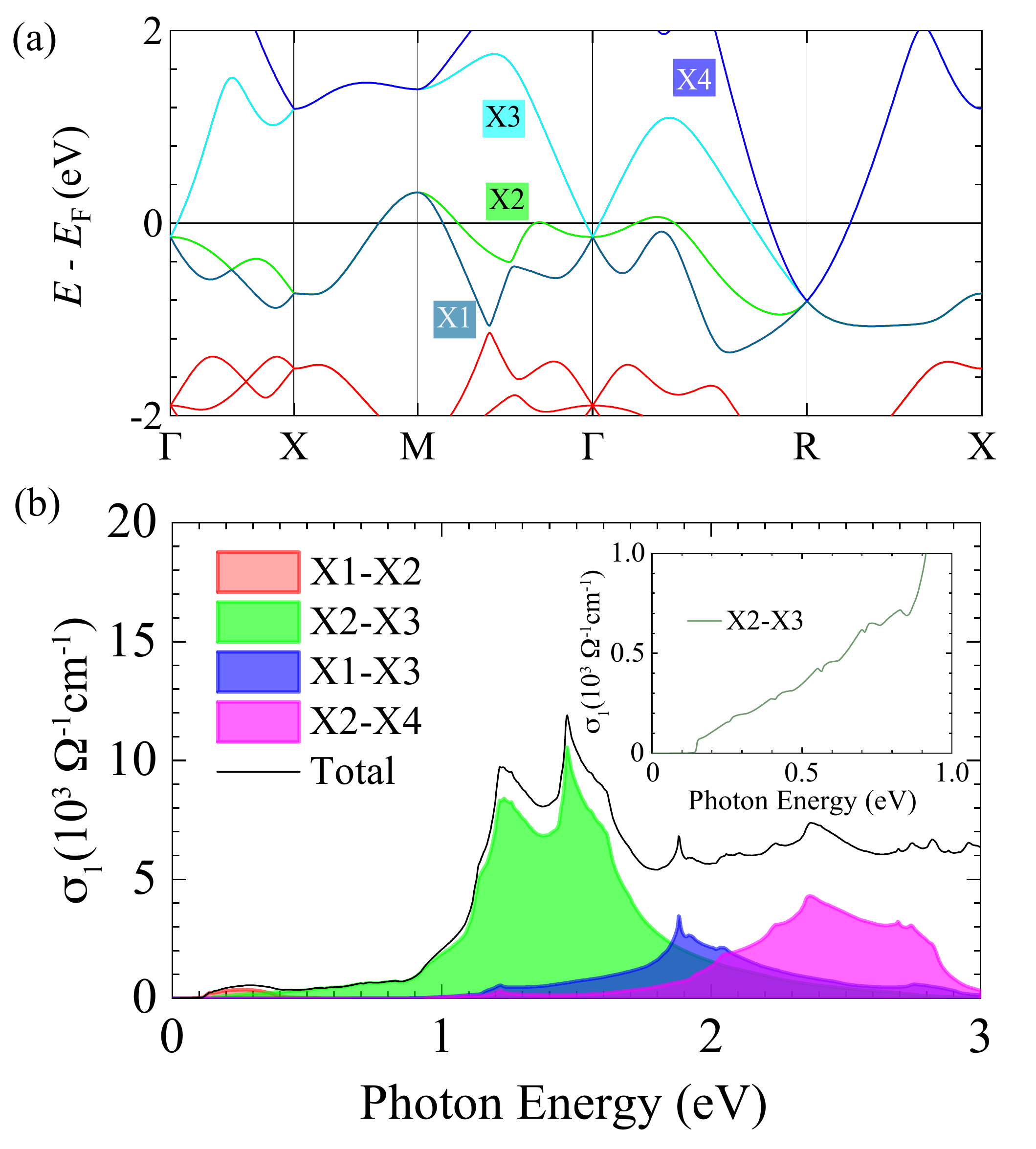}
	\caption{(a) Bulk band structure without SOC with band labelling as $\rm X1$, $\rm X2$, $\rm X3$, $\rm X4$. (b) Real part of calculated optical conductivity up to 3 eV photon energy, and inset represents only X2 $\rightarrow$ X3 band transitions.}
	\label{Fig5}
\end{figure}

Next we calculate the optical response function for  the photon polarization  along the $x$-axis. The optical conductivity is calculated for the interband transitions. The broadening parameter for the interband transitions is considered $0.1$~meV for weak disorders and the calculated total plasma frequency is $2.76$~eV. In the absence of SOC we identify bands as $\rm X1$, $\rm X2$, $\rm X3$, $\rm X4$ are crossing the Fermi energy level as noted in Fig~\ref{Fig5}(a). The optical conductivity contributions of specific band transitions are calculated such as X1 $\rightarrow$ X2, X1 $\rightarrow$ X3, X2 $\rightarrow$ X3, and X2 $\rightarrow$ X4. Different colours are used to show the isolated band transition in Fig~\ref{Fig5}(b). X1 $\rightarrow$ X2 broadband transition occurs in the lowest energy range, and X2 $\rightarrow$ X4 transition appears at the higher photon energy range.  Conductivity contribution of band transition X2 $\rightarrow$ X3 appears at photon energy of $\sim$ $0.14$ eV and is observed linear in a relatively large energy regime from $0.15$ to $0.85$ eV as represented in the green color of Fig~\ref{Fig5}(b) which is also  illustrated in the  inset of the same figure. This corroborates the robustness of large linear band relation ($\sim$~700 meV) in the electronic structure calculations. The summation of isolated conductivities is plotted in the black color render predominantly linear optical conductivity with the frequency from energy range $0.4$ to $0.8$~eV, suggesting the linear band transitions in $\rm PtAl$, as  observed for the linear band chiral topological systems $\rm RhSi$, $\rm CoSi$, $\rm PdGa$ ~\cite{polatkan2021optical, maulana2020optical, xu2020optical, maulana2021broadband}. 
The four-fold crossing point in PdGa is found far below the Fermi energy level therefore linear band feature in optical conductivity is observed in the high photon energy regime as compared to the PtAl~\cite{polatkan2021optical}.

\section*{CONCLUSIONS}
In summary, we have studied the transport and electronic properties of a single crystal of chiral quantum material PtAl. The Fermi surface mapping using the dHvA oscillations reveals the non-trivial nature of the $\alpha$ band that forms Weyl nodal fermion at the $\Gamma$ point. Higher-fold chiral fermions and robust surface states extended over the entire BZ are revealed. The existence of multifold chiral excitations in the bulk with longest surface Fermi arcs makes $\rm PtAl$ a fascinating material to explore topological phenomena and associated properties in experiments.  


%

\end{document}